\documentclass[5p]{elsarticle}     

\usepackage[T1]{fontenc}
\usepackage{times}

\usepackage{graphicx}
\usepackage{amssymb}
\usepackage{amsthm}
\usepackage{amsmath}
\usepackage{xypic}
\usepackage{url}
\usepackage[normalem]{ulem}

\usepackage{amsfonts}

\newtheorem{dfn}{Definition}
\newtheorem{theorem}{Theorem}
\newtheorem{example}{Example}

\newcommand{\mathset}[1]{\left\{#1\right\}}
\newcommand{\mathpset}[2]{\left\{#1\mid #2\right\}}

\newcommand{\qq}[1]{``#1''}

\newcommand{\switchtextoff}[1]{}

\DeclareMathOperator{\closure}{cl}

\begin{document}
\begin{frontmatter}    

\title{Applications of continuous functions in topological CAD data}


\author{Norbert Paul}
\address{Geodetic Institute of Karlsruhe (GIK), KIT, Englerstra{\ss}e 7, DE 76131 Karlsruhe\\
E-mail: norbert.paul@kit.edu}


\begin{abstract}
Most CAD or other spatial data models, in particular boundary representation models, are 
called \qq{topological} and represent spatial data by a structured collection of 
\qq{topological primitives} like edges, vertices, faces, and volumes. 
These then represent spatial objects in geo-information- (GIS) or CAD systems or in building 
information models (BIM). 
Volume objects may then either be represented by their 2D boundary or by a dedicated 
3D-element, the \qq{solid}. The latter may share common boundary elements with other 
solids, just as 2D-polygon topologies in GIS share common boundary edges. 

Despite the frequent reference to \qq{topology} in publications on spatial modelling the 
formal link between mathematical topology and these \qq{topological} models is hardly described 
in the literature. Such link, for example, cannot be established by 
the often cited nine-intersections model which is too elementary for that purpose. 
Mathematically, the link between spatial data and the modelled \qq{real world} entities 
is established by a chain of \emph{continuous functions}---a very important topological notion, 
yet often overlooked by spatial data modellers. 
This article investigates how spatial data can actually be considered topological spaces, 
how continuous functions between them are defined, and how CAD systems can make use of them. 
Having found examples of applications of continuity in CAD data models 
it turns out that of continuity has much practical relevance for CAD systems. 
\end{abstract} 

\begin{keyword}
Topological Data Modelling\sep 
Mathematical Topology\sep 
Spatial Consistency Rules\sep 
Topological Constructions
\end{keyword}

\end{frontmatter}    

\section{Introduction}\label{sec:intro}

Spatial data modelling is an important task in almost all engineering applications. 
Current CAD systems provide spatial models for engineering, and standards have been 
established to exchange such data between different applications such as STEP for generic 
engineering and, say, the Industry Foundation Classes (IFC) \cite{IFC2x3} for the building 
and construction domain. 
However, the overall objective of these standardisation efforts towards an integrated 
multi-user and model-based design environment has not yet been reached \cite{Nepal2012904}. 

Topology denotes the connectivity between, for example, the rooms of a building and their 
links like doors, or walls. 
It also describes the connectivity among these elements. 
This connectivity can be considered at different levels of detail: 
A wall may connect two room volumes but it may also be considered a volume connected to 
each room by its boundary common surface. 
The \qq{part\_of}-relation between these wall-parts and the wall object is already 
an example of a continuous function.
Now, despite the frequent use of the vocabulary \qq{topology} in spatial data modelling 
publications, only few of them actually use topological methods 
\cite{Boltcheva:Homology,Raghothama:CRT}. 
Whereas in mathematics continuous functions are an indispensable part of topology and omitting 
them is just unthinkable, publications on \qq{spatial data} and \qq{topology} are much more 
frequent than those on \qq{spatial data} and either \qq{continuous function} or \qq{continuous map}.
Topology provides an extremely useful mathematical framework for CAD data management 
for which it might be as important as predicate logic and set theory is for data 
modelling in general. Note that \cite{Vella:GraphTheoryIsTopology} even proposes topology as a 
generalisation of graph theory. 

This article presents the result of an investigation to locate the open sets, topological 
spaces, and continuous functions from point set topology within CAD data instances, how 
continuity can be expressed in terms of the data representations of these topological spaces 
and which importance continuous functions have therein. 
Section~\ref{sec:basics} it introduces the most common 
generic layout of a topological spatial data model and then introduces \qq{topology} and 
\qq{continuous function}. Then an idealised spatial modelling \qq{process} 
transfers \qq{the} topology of the embedding space to the spatial data entities.  
Section~\ref{sec:data-topology} identifies \qq{topology} and \qq{continuous functions} 
within such data models and thus shows that the \qq{topology} of a 
CAD model is indeed a mathematical topology which turns the model itself into a 
space. Such data is linked to the topology of the embedding space $\mathbb{R}^3$ by 
continuous functions which form a \qq{bucket chain} that passes the topology 
from $\mathbb{R}^3$  into the data model. 
Continuity decides if the spatial model is topologically \qq{correct} and is therefore 
an interesting consistency constraint. 
Sections~\ref{sec:applications} and~\ref{sec:applications2} present other applications 
of continuity. 


As this article focuses on point-set topology other important aspects of data modelling like, 
for example, semantics, efficency, uncertainty, or object-orientation may not be mentioned here. 
This does not question their importance but rather means that they are outside the article's 
scope. 

\section{Related work}

In contemporary mathematics \emph{Topology} is considered one of the most fundamental discipline. 
It also plays an important r\^ole in spatial data modelling. But many of its possibilities, 
like the important \emph{topological constructions} \cite[Ch.\ IV]{Viro:Textbook},
are currently left unexploited. 

\qq{Topological spaces} are, like \qq{vector spaces}, abstractions of the 
geometrical structures of the Euclidean space $\mathbb{R}^3$. 
After linear algebra has dropped its restriction to $\mathbb{R}^3$ due to the work of 
Cayley \cite{Tait:Obituary} it conquered a wide field of applications like 
finite element computation, or simulation. 
Topology, too, is not restricted to \qq{spatial} models in $\mathbb{R}^n$ the intuitive 
sense but also covers abstract \emph{spatial} models \cite{BrahaReich:Design}. 

Alexandrov spaces \cite{Alexandroff} are a particularly important class of 
topological spaces for computational topology \cite{Arenas:AlexandroffSpaces} 
and CAD data modelling. 
They also generalise undirected graphs which makes topology a sensible supplement of graph 
theory \cite{Vella:GraphTheoryIsTopology}. 

The \emph{relational model} \cite{Codd:RelationalModel} was the first data model based on 
a strictly formal mathematical theory. Its query \qq{language} is the 
\emph{relational algebra}---a definition of the formal semantics of elementary 
query operators with a deliberate omission of any language syntax 
\cite[p.\ 21]{Codd:RelationalModel}. A characteristic of this model is its 
stunning \emph{simplicity} compared to the database models that have existed so far. 
This simplicity enabled the development of complex applications, and 
a general database-design theory \cite{Codd:DataModels}. 
In contrast, \cite{Amor:BIM2007} identifies unnecessary complexity in the IFC schema.  
Note that \cite{Watson:Challenges} states that from the \qq{progressively increasing complexity} 
of the future building and construction domain new challenges of BIM will emerge. 


The STEP (STandard for the Exchange of Product model data), standardises the exchange of 
product models \cite{Aug95, Eas99} and is a basis for CAD models or the IFC\@. 
These models combine spatial and domain-specific information and usually are very complex 
\cite{Amor:BIM2007, Eas99}. But even though the IFC cover a broad spectrum of 
topological modelling elements \cite{Paul-2010} surprisingly they still seem to lack topological 
functionality needed: At least Rank e.a.\ define their own model to 
access topological information of an IFC model \cite{SPP1103:RankEtAl}. 
A very important topological property of a space, or of spatial data, is its 
\emph{homology}. Boltcheva e.a.\ \cite{Boltcheva:Homology} present an efficient algorithm to 
compute simplicial homology. 

An example of topological spaces beyond $\mathbb{R}^3$ in the context of relational databases 
is the theory of \emph{acyclic database schemata} which converts a database schema into a 
topological space and tests it for so-called $\alpha$-cycles. 
This topological property severely affects the efficiency of some database query 
operations \cite{Fagin:acyclicdatabase}.

\section{Basic notions}\label{sec:basics}

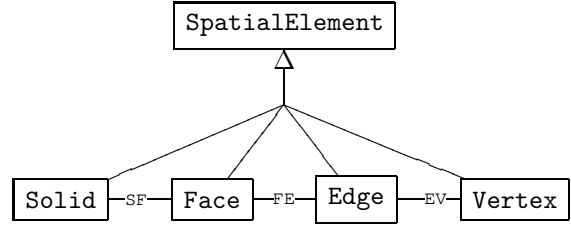
\begin{figure}[h]
$$
\newcommand{\bdwidth}{9pt}
\xymatrix{
              &\ar@{}[r]|*+<\bdwidth>[F]{\tt SpatialElement}="top" &\\ 
              &\ar@{}[r]|{}="head"\ar@{-}"head";"top"|(.56){\displaystyle{\Delta}}&&\\
       *+<\bdwidth>[F]{\tt Solid}\ar@{-}"head" \ar@{-}[r]|{\texttt{SF}} & 
       *+<\bdwidth>[F]{\tt Face}\ar@{-}"head"  \ar@{-}[r] |{\texttt{FE}}& 
       *+<\bdwidth>[F]{\tt Edge} \ar@{-}"head" \ar@{-}[r] |{\texttt{EV}}& 
       *+<\bdwidth>[F]{\tt Vertex}\ar@{-}"head"
}
$$
\caption{The classical layout of a \qq{topological} data model---a sequence of classes of spatial 
entities connected to the typical \qq{cell chain}, or complex, by incidence relations \texttt{SF}, 
\texttt{FE}, and \texttt{EV}. These induce a relation on \texttt{SpatialElement}.}
\label{fig:spatialUML}
\end{figure}

Figure~\ref{fig:spatialUML} introduces a class diagram of an abstract topological 3D 
data structure which is frequently used up to minor modifications.
This model presents the classical sequence of 3-, 2-, 1-, 0-dimensional entity types 
where two consecutive entity types are connected by an association into the typical 
\qq{chain} structure. The length of the sequence corresponds with the dimension of the 
model and may vary: 2D GIS \qq{polygon topologies} have no \texttt{Solid} class and a 
4D-application could append a \texttt{HyperSolid} class to the left. 

A vertex usually represents a point in $\mathbb{R}^n$ modelled by storing its coordinates. 
An edge represents a path from a starting vertex to an ending vertex and the data 
structure relates each edge entity with two vertex entities. 
Some edges may cycle around a face which then is modelled by a face entity 
related to its boundary edges by an $n:m$ relationship. 
Faces may have holes which then partition its boundary edges into components. 
Some data structures express this partitioning explicitly by providing \texttt{Loops}. 
We will later see that loops are redundant and can be computed by, say,  
using \cite{Boltcheva:Homology}. 
Curiously, arrangements for edges with holes are uncommon. These would then have more 
than two boundary vertices and will be discussed later. 
When faces surround a solid that solid is represented by a \texttt{Solid} entity 
related to its faces by a Solid--Face association. 

\subsection{Topology}

After the following introduction of the topological structure of the embedding space 
it will be obvious that this structure also applies to spatial data. 

The property of the embedding space which leads to its topology is the 
definition of a \emph{distance} for every pair of points in a space:
\begin{dfn}[Metric space]
Let $X$ be a set and $d:X\times X\to \mathbb{R}$ be a function that maps 
two points $a$ and $b$ to a real value $d(a,b)$. Then $d$ is called a 
\emph{metric} on $X$, iff $d$ satisfies the following properties for all 
$a$, $b$, and $c$ in $X$:
\begin{enumerate}
\item $d(a,b)\ge 0$ and $d(a,b)= 0 \Leftrightarrow a=b$ (non-negativity),
\item $d(a,b)= d(b,a)$ (symmetry),
\item $d(a,b)+d(b,c) \ge d(a,c)$ (triangle inequality).
\end{enumerate}
When $d$ is a metric for $X$ then the pair $(X,d)$ is called a \emph{metric space}.
\end{dfn}
The classical example of a metric space is the 
$n$-di\-men\-sio\-nal 
real space $\mathbb{R}^n$ with its Euclidean metric 
$$
  d(a,b) = \sqrt{(a_1 - b_1)^2 + \dotsc (a_n - b_n)^2}\enspace.
$$
Each metric allows a generalisation of open intervals: 
\begin{dfn}[Open ball]
Let $(X,d)$ be a metric space, $c\in X$ an arbitrary (centre) point in $X$ and 
$r$ an arbitrary positive real number $0 < r\in \mathbb{R}$ (radius).  
Then the set $B(c,r) := \mathpset{b\in X}{d(b,c)<r}$ is called an \emph{open ball} 
in $(X,d)$.
\end{dfn}
The open balls in $\mathbb{R}$ are the \qq{open intervals} and 
$B(c,r) \in \mathbb{R}$ is then usually written as $(c-r,c+r)$. 
From now on $\bigcup\mathcal{I}$ will be a shorthand notation for $\bigcup_{I\in\mathcal{I}}I$. 
Every union $\bigcup\mathcal{I}$ of a set $\mathcal{I}$ of open balls will be called an 
\emph{open} set in $(X,d)$. 
The set $\mathcal{T}_{\mathbb{R}^n}$ of all open sets in $\mathbb{R}^n$ has three fundamental 
properties which are the defining properties (axioms) of a topology:
\begin{dfn}[Topology, topological space]
Let $X$ be a set. A set $\mathcal{T}_X$ of subsets of $X$ is called a \emph{topology} 
for $X$ if it satisfies the following three properties:
\begin{enumerate}
\item $\mathbb{X}$ and $\emptyset$ are elements of $\mathcal{T}_X$.
\item The union $\bigcup\mathcal{S}$ of every subset $\mathcal{S}$ of $\mathcal{T}_X$ is an 
      element of $\mathcal{T}_X$.
\item The intersection $A\cap B$ of every two elements $A,B$ of $\mathcal{T}_X$ is an 
      element of $\mathcal{T}_X$.
\end{enumerate}
Then the pair $\underline{X} := (X,\mathcal{T}_X)$ is called a \emph{topological space}.
Every element of $\mathcal{T}_X$ is called an \emph{open set} in $\underline{X}$, and 
an element of $X$ is called a \emph{point} in $\underline{X}$.
\end{dfn}
So any element of any set may be a \qq{point} and a point set can have a many topologies. 
The above topology $\mathcal{T}_{\mathbb{R}^n}$ is called the \qq{natural topology} 
of $\mathbb{R}^n$. 
But the power set $\mathcal{P}(X)$---the set of all subsets of 
$X$---is also a topology for $X$ which gives the so-called 
\emph{discrete space} $(X,\mathcal{P}(X))$. 
So there is no \qq{topological property} of an element by itself. 
This \qq{pluggability} of topologies in mathematics is usually missing in spatial data models 
where topological properties are often hard-coded into the elements. 
But because an element may exist in different spaces its topological properties, say, its 
dimension, may vary therein. 

\begin{sloppy} 
\begin{example}[Room connectivity]\label{exa:connectivity} 
Such change of dimension occurs in the \qq{room connectivity graph} of a 
building (e.g.\ \cite{Jensen:Indoor}): 
A building model is a topological space which contains 
three-dimensional room objects and two-dimensional connecting doors and other 
\qq{virtual} passages, say, a door frame without a door. 
Later this article will introduce the topological subspace. 
One subspace of a building consists only of its rooms, doors and passages 
and has dimension one: the rooms therein are 1D and the doors are 
of dimension zero. 
The \qq{connectivity graph} is merely its \qq{dual space} where the dimensions 
of the elements are inverted. 
Within this space the passages and doors become one-dimensional edges of the 
connectivity graph each connecting two rooms as its zero-dimensional vertices. 
This \emph{point set} topological duality is intimately related to the Poincar\'e 
duality from \emph{algebraic} topology. 
\end{example}
\end{sloppy} 

\subsection{Continuous functions}
It is one of the achievements of topology to establish a 
ge\-ne\-ric 
definition of 
\qq{continuous function}. 
In calculus real-valued functions $f:\mathbb{R}\to\mathbb{R}$ are called \emph{continuous} 
iff (if and only if) \qq{small changes} of the function argument $x$ only lead to 
\qq{small changes} in the functional value $f(x)$. Its formal definition with the classical 
$\epsilon$-s and $\delta$-s---distances that represent these small changes---will be omitted 
here. 
A generalised \qq{continuity} for data structures with a finite number of 
elements cannot use a metric and so \qq{small change} must be adapted accordingly:
Let $S\subseteq X$ be a set of points of a space $(X,\mathcal{T}_X)$. 
A point $p\in X$ is said to be \emph{close} to $S$ (within $(X,\mathcal{T}_X)$) if every open 
set that contains $p$ intersects $S$. 
The set of all points close to $S$ is denoted by $\closure S$ the 
\emph{closure} of $S$. For example, the boundary points of an open interval are not 
members but close points of the interval, hence $\closure (a,b) = [a,b]$. 
Using the notion of \qq{close} instead of \qq{small change} gives a purely topological and 
very illustrative definition of continuous functions:

\begin{dfn}[Continuous function]
Let $\underline{X}=(X,\mathcal{T}_X)$ and $\underline{Y}= (Y,\mathcal{T}_Y)$ be two 
topological spaces. 
A function $f:X\to Y$ is called \emph{continuous} iff for every point $p\in X$ and every set 
$A\subseteq X$ such that $p$ is close to $A$ in $\underline{X}$ the image point $f(p)$ is close 
to the image set $f[A]$ in $\underline{Y}$. 
The fact that $f$ is continuous will be denoted by $f:(X,\mathcal{T}_X)\to(Y,\mathcal{T}_Y)$.
\end{dfn}

So continuous functions respect the \qq{being-close} property within a space. The definition 
only depends on the topology and forgets any possibly underlying metric. 
It is equivalent to the calculus definition with the natural topology but it can be applied 
to every topological space. The above definition is also equivalent to the usual topological 
definition of continuity:

\begin{theorem}[Continuous function]
Let $\underline{X}=(X,\mathcal{T}_X)$ and $\underline{Y}= (Y,\mathcal{T}_Y)$ be two 
topological spaces. 
A function $f:X\to Y$ is \emph{continuous} iff for every open set $U\in\mathcal{T}_Y$ 
the pre-image $f^{-1}[U]:=\mathpset{x\in X}{f(x)\in U}$ is open in $\underline{X}$.
\end{theorem}

Now a spatial data modeller must find a data representation of spatial objects in 
a computer. The following idealised spatial modelling process transfers a physical 
spatial object into a topological data model and shows that this process is connected 
by a chain of continuous functions.

\section{An idealised spatial modelling process}

The idealised spatial modelling \qq{process} does a step-by-step transfer of an 
assumed \qq{real-world} spatial object from its embedding space $\mathbb{R}^3$ into its 
data representation. 

These steps are: 
First, selecting the (infinite) set of points of the embedding space which are 
\qq{occupied} by some element of the given object. 
Then identifying each point set that belongs to a spatial element of that object 
by \emph{partitioning} the whole point set into a finite number of parts. 
This creates a topology for these parts, which is an Alexandrov-topology \cite{Alexandroff}
and hence has a representation as a preordered set, usually a partially ordered set (poset). 
Finally, for each such element a data entity to represent the corresponding 
element will be stored together with a relation that represents the poset: 
Every finite partial order is the transitive and reflexive closure of a directed 
acyclic graph (DAG). 

This section shows that these steps are connected by continuous functions thus 
exposing continuity as \emph{the} key consistency rule for \qq{topological correctness} 
of spatial data.  

\subsection{Step 1: Carving out the overall shape}

\begin{figure}[b]
$$\xymatrix@=16pt@M=2pt{
  & \bullet \ar@{-}[rr]& & \bullet\\
\bullet\ar@{-}[rr]\ar@{-}[ru]& & \bullet\ar@{-}[ru]\\
  & \bullet \ar@{.}[rr]\ar@{.}[uu]& & \bullet\ar@{-}[uu]\\
\bullet\ar@{-}[rr]\ar@{-}[uu]\ar@{.}[ru]& & \bullet\ar@{-}[ru]\ar@{-}[uu]\\
  & \bullet \ar@{.}[rr]\ar@{.}[uu]& & \bullet\ar@{-}[uu]\\
\bullet\ar@{-}[rr]\ar@{-}[uu]\ar@{.}[ru]& & \bullet\ar@{-}[ru]\ar@{-}[uu]\\
}$$
\caption{The \qq{two storey house} with two volumes ($3$-manifolds). Its walls, roof, and slabs 
are $2$-manifolds, the edges are $1$-manifolds and the vertices are $0$-manifolds.} 
\label{fig:house}
\end{figure}
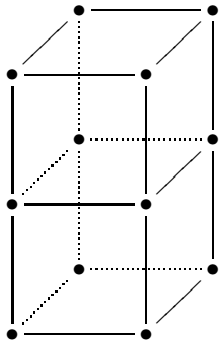

The set $\mathbb{R}^3$ of the three-dimensional points with its natural topology represents the 
physical space containing the object in consideration. 
For simplicity just two unit cubes $[0,1]^3$ and $[0,1]^2\times[1,2]$ piled atop are considered. 
This gives a simple specimen of a two-storey \qq{house} as depicted in Figure~\ref{fig:house}. 

The notation $[a,b]$ denotes $\mathpset{x\in\mathbb{R}}{a\le x\le b}$, the \emph{closed} interval
with $a$ and $b$ as elements. The two cubes are Cartesian products of such intervals. 

The selection $H$ of the points occupied by the object gives   
$H = [0,1]\times[0,1]\times[0,2]$ a prism over a square in the 
$x,y$-plane. Selecting a set of points from a topological space also returns a 
topology. So the selection becomes a topological subspace: 
\begin{dfn}(Subspace)
Let $(X,\mathcal{T})$ be a topological space and let $H$ be a subset of $X$. 
Then the set 
$$\mathcal{T}|_H := \mathpset{H \cap U}{U \in \mathcal{T}}$$ 
is a topology for $H$. 
The topological space $(H,\mathcal{T}|_H)$ is called the \emph{subspace} of $H$ 
in $(X,\mathcal{T})$. 
\end{dfn}
Note that selection is also a relational query operator. 
The continuous function involved in this construction is 
the \emph{inclusion function} 
$$
   i:H\hookrightarrow \mathbb{R}^3, h \mapsto i(h) := h
$$
which relates the domain \emph{set} $H$ with the range \emph{space} 
$\mathbb{R}^3$. 
It is the \qq{engine} that drags the topology from the space to the set: 
The subspace topology is the unique minimal topology that turns the inclusion 
function $i$ into a continuous function. 

\begin{sloppy} 
This \qq{handing over topologies} now takes place along a \qq{bucket chain} of functions that connects 
the real-world space $\mathbb{R}^3$ containing the object with the data model instance in the 
computer. These are all called \qq{topological constructions} and covered, for example, in topology 
text-books like \cite[Ch.\ IV]{Viro:Textbook}. 
\end{sloppy} 

\subsection{Step 2: Specifying the features}

After turning $H$ into the space $(H,\mathcal{T}|_H)$ all is ready for the next modelling step: 
Specify the features the model is composed of. 
The example on Figure~\ref{fig:house} partitions $H$ into 12 vertices, 20 edges, 
11 faces, and 2 solids. 
Here, for simplicity all features are considered manifolds. 
It will later be diccussed if a manifold assumption can be realistic. 
\qq{Fat features} would do, too, and walls of non-zero 
thickness may be connected together by, say, columns the ends of which run into 
connecting node constructions represented by volume objects. 
But all features must be non-empty disjoint point sets. 
For example, a point at such a fat wall's surface should belong to the wall and not to a room. 
The \qq{features} are non-empty subsets of $H$ that together cover $H$. 
A disjoint covering of a set $H$ by non-empty subsets is called a \emph{partitioning} of $H$ 
and usually denoted by $H/_{\!\sim}$. Each point $p$ in $H$ has exactly one corresponding 
point set $[p]\in H/_{\!\sim}$ with $p\in[p]$ which gives another function 
$$
  \pi:H\to H/_{\!\sim}, p \mapsto \pi(p) := [p]
$$
the \emph{natural projection} from a single point to its part. 
The symbol $\sim$ denotes the induced equivalence relation: Two points are equivalent, iff 
they belong to the same \qq{feature}. 

Again, a function connects a space $(H,\mathcal{T}|_H)$ with a yet unstructured 
set (of sets) $H/_{\!\sim}$ which now also gets a topology: 

\begin{dfn}[Quotient topology]
Let $(X,\mathcal{T})$ be a topological space and $X/_{\!\sim}$ be a partitioning of $X$. 
Then the set $\mathcal{T}/_{\!\sim} := \mathpset{A\subset X/_{\!\sim}}{\bigcup A \in \mathcal{T}}$ 
is called the \emph{quotient topology} of $\mathcal{T}$ with respect to $\sim$. 
The topological space 
$$
  (X,\mathcal{T})/_{\!\sim} := (X/_{\!\sim},\mathcal{T}/_{\!\sim})
$$ 
is called the \emph{quotient space} of $(X,\mathcal{T})$ with respect to $\sim$.
\end{dfn}

The quotient topology is the minimal topology such that $\pi$ is continuous. 
It has taken us one step closer to the spatial data model where each data entity 
represents a feature of $H/_{\!\sim}$. 
Note the duality to the first step: Here the topology is passed from a domain \emph{space} 
to a range \emph{set} by $\pi$ whereas the inclusion function passed it from the range to the 
domain. 
Accordingly, the quotient topology is the \qq{biggest} topology such that the function 
$\pi$ stays continuous. 
So when $i$ was the engine that pulled the topology from $\mathbb{R}^3$ into $H$ then $\pi$ 
is a dual engine that shoves it on to $X/_{\!\sim}$ thus making itself continuous. 

Features may be partitioned again, say, by grouping manifolds into fat features. 
This then gives a \qq{quotient of a quotient} and turns $\pi$ into a \qq{part\_of} relation. 
Repeating this gives a formal model of different levels of detail.

\subsection{Step 3: Relations and topologies}

Each topological space $(X,\mathcal{T}_X)$ has an associated preorder relation 
$\succeq$ on $X$ defined as $b \succeq a \Leftrightarrow a \in \closure\mathset{b}$ which 
is called the \emph{specialisation preorder} of $(X,\mathcal{T}_X)$. 
Conversely, each relation $R$ on a set $X$ defines a topology 
\begin{align}\label{eqn:relationaltopology}
  \mathcal{T}(R) 
    := \mathpset{A \subseteq X}{\forall (a,b)\in R : b\in A \Rightarrow a\in A}\enspace.
\end{align}
The preorder $R^{*}$, the transitive and reflexive closure of $R$, is the specialisation preorder 
of $\mathcal{T}(R)$. 
Note that bigger relations give smaller topologies: 
$R\subseteq S\Rightarrow \mathcal{T}(S)\subseteq \mathcal{T}(R)$. 
We will hence call a relation $R$ \emph{finer} than $S$ if $R\subseteq S^{*}$ holds. 
Then $S$ is \emph{coarser} than $R$.
Eqn.~\ref{eqn:relationaltopology} indeed satisfies the three axioms and thus denotes a topology: 
The first is easy to see and the proof of the second and the third are similar. So only the 
second axiom will be proven here leaving the third axiom to the reader: 

\begin{proof}
Let $\mathcal{A}\subset\mathcal{T}(R)$ be an arbitrary set of open sets. 
Then each satisfies the above property. 
Assume $a\,R\,b$ and $b\in\bigcup\mathcal{A}$ hold for an arbitrary $b$. 
By $b\in\bigcup\mathcal{A}$ there exists a set $A\in\mathcal{A}$ which contains $b$. 
By $a\,R\,b$ the set $A$ also contains $a$ because it is open. 
Therefore $a$ is in $\bigcup\mathcal{A}$. 
Hence $\bigcup\mathcal{A}$ is also open. 
\end{proof}

The similar proof of the third axiom shows that an even stronger property holds: 
Every set of open sets has an open intersection. The finiteness restriction of 
axiom three can be dropped which characterises \emph{Alexandrov topologies} \cite{Alexandroff}:

\begin{dfn}[Alexandrov topology]
Let $\underline{X} = (X,\mathcal{T})$ be a topological space. 
Then $\mathcal{T}$ is called an \emph{Alexandrov topology}, iff 
the intersection $\bigcap\mathcal{S}$ of every subset 
$\mathcal{S}$ of $\mathcal{T}$ is an element of $\mathcal{T}$. 
Then $\underline{X}$ is called an \emph{Alexandrov topological space}. 
\end{dfn}

Alexandrov stated, already in 1937, that these spaces (with a restriction called $T_0$) 
are essentially the same as the partially ordered sets $(X,R^{*})$. 
There are two important properties of Alexandrov spaces: 
First, a space is an Alexandrov space, iff it can be generated by a relation, and, 
second, every finite topological space is an Alexandrov space. 
Interestingly, Sorkin even proposes this topology in theoretical physics 
as a model for space-time itself because space-time might have an \qq{underlying discreteness} at 
quantum-level where the assumptions of a $4$-manifold space-time fail \cite{Sorkin:FinTop}. 

This poset vs.\ topology equivalence immediately leads to an extremely simple data model 
which can represent \emph{every} topology for a finite set: 

\begin{dfn}[Topological data type]
A simple directed graph $(X,R)$ with a set $X$ of elements and a binary 
relation $R\subseteq X\times X$ is called a \emph{topological data type}. 
The topological space $(X,\mathcal{T}(R))$ is the called the topological 
space \emph{represented by} $(X,R)$. 
\end{dfn}

$R$ need not be reflexive or transitive but it will be assumed acyclic in 
this article (to guarantee above mentioned $T_0$ restriction) 
and will be called an \emph{incidence relation} of $\mathcal{T}(R)$ because 
special cases of topological data types are called \qq{incidence graphs} 
\cite[p.\ 395]{Brisson:TopologyOrder}.

The above partitioning of the infinite set $H$ into the finite set $H/_{\!\sim}$ leads to 
a finite topology $\mathcal{T}/_{\!\sim}$ which then has a minimal incidence relation $Q$ on 
$H/_{\!\sim}$ such that  $\mathcal{T}/_{\!\sim} = \mathcal{T}(Q)$ with $Q^{*}$ as its specialisation 
preorder. So the final modelling step merely consists in storing a data representation of the 
finite graph $(H/_{\!\sim},Q)$.

\subsection{Step 4: Finally, the data representation}

In the final step each feature in $H/_{\!\sim}$ will be represented by a data entity. 
For example, each vertex $(r_x,r_y,r_z)\in \mathbb{R}^3$ gets a unique vertex identifier 
$n$ and an according entity, say 
$$
   \texttt{Vertex(id:$n$, x:$r_x$, y:$r_y$ ,z:$r_z$)}\enspace,
$$
is stored in some data structure. 
Every edge gets an identifier $\mathit{eid}$ and is stored as 
\texttt{Edge(id:$\mathit{eid}$)}. For each vertex $m$ close to $\mathit{eid}$ the association 
\texttt{(Edge.id:$eid$,Vertex.id:$m$)} has to be stored, too.
Then face entities are stored in \texttt{Face} and, accordingly their \qq{close} 
relation is stored as a \texttt{Face}--\texttt{Edge} association. Finally, the 
solids and their corresponding \texttt{Solid}--\texttt{Face} relationships are to be stored. 
The above example partitioning of $H$ is \qq{docile} in the sense that the \qq{bounded by} 
relation $Q$ on $H/_{\!\sim}$ can, in fact, be represented by the three 
associations below \texttt{Spa\-tial\-Ele\-ment} but not every partitioning of $H$ has 
this property. 

\section{The continuity on spatial data}\label{sec:data-topology}

The above mentioned associations define a relation named 
\qq{bounded by} on the common superclass 
\texttt{Spa\-tial\-Ele\-ment}. The members of this class will here be denoted by the set $X$ 
and the incidence relations by one relation $R\subseteq X\times X$. 
Note that from $(X,R)$ the dimension of each element 
$x\in X$ can be computed: 
It is the maximal number $n$ of elements $x_n,\dotsc,x_1\in X$ 
that can be appended to $x$ in a chain $x\,R\,x_n\,R\dotsc R\,x_1$. 
So by unifying all classes into $X$ no information gets lost. 
Vella, too, proposes the union of edges and vertices of a graph into a set 
$X$ \cite{Vella:GraphTheoryIsTopology}. 
Also each chain $x_a\,R^{+}\dotsc\,R^{+}\,x_b$ of length $m$ can be considered an abstract 
$m$-simplex $\langle x_a,\dotsc, x_b\rangle$ of an abstract simplicial complex. This has the 
same homology as $(X,\mathcal{T}(R))$. So the algorithm of \cite{Boltcheva:Homology} can find 
all loops and shells in $(X,R)$. As mentioned above, explicit loops and shells are 
redundant. 

The function $\sigma:(H/_{\!\sim},\mathcal{T}/_{\!\sim})\to(X,\mathcal{T}(R))$, which maps each 
feature in $H/_{\!\sim}$ to its representing data entity in $X$, is continuous when each pair 
$([a],[b])$ of close features has an image pair $(\sigma([a]),\sigma([b]))$ in $R^{*}$ of 
close data entities. The following statement generalises this:

\begin{sloppy} 
\begin{theorem}[Continuity on topological data types]
Let $(X,R)$ and $(Y,S)$ be two topological data types, and let ${f:X\to Y}$ be a function 
from $X$ to $Y$. Then $f$ is continuous from $(X,\mathcal{T}(R))$ to $(Y,\mathcal{T}(S))$ 
iff the image pair $(f(a),f(b))$ of every pair $(a,b)\in R$ is an element of 
$S^{*}$. 
Continuity between topological data types will be denoted by $f:(X,R)\to(Y,S)$.
\end{theorem}
\end{sloppy} 

The proof can be found in \cite[Thm.\ 5.6]{BradleyPaul}. 
The representation $(X,R)$ has the flexibility to also store spaces that result from a 
\qq{non-docile} partitioning which may result in a relation $R$ that cannot be spread into a 
chain of, say, three associations. It also allows dynamic change of dimension at run-time.

As $\sigma$ simply carries an incidence relation $Q$ of $(H/_{\!\sim},\mathcal{T}/_{\!\sim})$ to 
$R^{*}$ it is continuous. 
But $\sigma$ is a bijection and the inverse function 
$$
   \sigma^{-1}:(X,\mathcal{T}(R))\to(H/_{\!\sim},\mathcal{T}/_{\!\sim})
$$ 
is continuous, too. 
The features space $(H/_{\!\sim},\mathcal{T}/_{\!\sim})$ and the data space $(X,\mathcal{T}(R))$ 
created by $(X,R)$ are topologically 
equi\-valent 
or \qq{homeomorphic}. So $(X,R)$ is a topologically 
consistent representation of $(H/_{\!\sim},\mathcal{T}/_{\!\sim})$.

\subsection{The bucket chain of continuous functions}

As presented above instances of spatial data are topological spaces and a chain of continuous 
functions 
$$
  \mathbb{R}^3 \ \stackrel{i}{\hookleftarrow}\
             H \ \stackrel{\pi}{\longrightarrow}\ 
             H/_{\!\sim}\ \stackrel{\sigma}{\longrightarrow}\ X
$$
links the entities $X$ of a topological data type $(X,R)$ with the \qq{real world} topological 
space $\mathbb{R}^3$ which contains the object. These functions define the relation $R$ in the 
data type by passing the topology from left to right. 

\subsection{Homeomorphism of spatial data} 

Continuity immediately determines when two instances of spatial data are 
topologically equivalent or \qq{homeomorphic} just as with every topological space. 
The definition of homeomorphism needs the \emph{identical function} on $(X,R)$ 
$$
  \operatorname{id}_{(X,R)}:(X,R)\to(X,R), x\mapsto x\enspace.
$$
This special case of a continuous inclusion function is trivial but not 
less important than other mathematical trivialities like the number zero or the empty set. 

\begin{sloppy} 
\begin{dfn}[Spatial data homeomorphism] Let $(X,R)$ and $(Y,S)$ be instances of topological 
data types and let 
${f:(X,R)\to(Y,S)}$ 
and 
${g:(Y,S)\to(X,R)}$ 
be two continuous functions. 
Then $f$ is called a \emph{homeomorphism} iff the two function compositions 
$f\circ g$ and $g\circ f$, with $f\circ g(x)=f(g(x))$ and $g\circ f(x)=g(f(x))$, 
are the identical functions on $(X,R)$ and $(Y,S)$. 
Two instances of topological data types are called \emph{homeomorphic}, iff a homeomorphism 
between them exists.
\end{dfn}
\end{sloppy} 

It is known to be a hard problem to decide if two given topological data types are homeomorphic: 
\cite{PaulDiss} shows that the homeomorphism problem is graph 
isomorphism complete (and \cite{BrettoEtAl:CompatibleTopologies} did so two years earlier). 
The graph isomorphism problem defines a complexity class GI which is supposed to be a bit harder 
than P but not as hard as NP \cite{Hoffmann:GI}. 

\section{Applications of continuous functions}\label{sec:applications}

Identifying continuity of functions as the formal link between real world spatial objects and 
their topological data representation might be considered a merely abstract result of no 
practical relevance. 
But another practical application of continuity has already been mentioned here: 
Consistently linking different levels of detail (LoDs)---from the finest, say,  
\qq{geometric representation} to the corsest \qq{element connectivity}---by 
a chain of continuous functions, say, \qq{part\_of} in a CAD model. 
This section presents other useful applications of continuity. 

It studies continuous functions between two instances of topological 
data types in an abstract manner, ignoring any \qq{outside} topology of a surrounding world or 
any other semantics. 
Having only one universal structure for spatial entities gives the flexibility to map, say, a 
column volume and its surface at a higher LoD to a simple edge representation of 
said column at a lower LoD, or, short, to map a mix of 0D, 1D, 2D, and 3D-entities to 
an $n$D-entity of arbitrary dimension. So continuity between topological data types can become 
a consistency rule.   

Also the topological constructions, like the above presented subspace and 
quotient space, cover a relationally complete set of topological query operators. 
These take input spaces and produce output spaces similar to relational algebra 
which turns input relations to output relations. 
The input and output spaces are always linked together by continuous functions which 
define the topology of the query result.

\subsection{Continuous foreign key}

CityGML's LoD-references currently permit the coarser 
object to be of completely different 
shape and at a completely different location than its finer representation 
at a higher LoD \cite[p.\ 314]{OosteromStoter:5D}. 
But objects at different LoDs are topological spaces. So the association between detailed 
and coarser representation could be restricted by a consistency rule that ensures that the 
foreign key represents a continuous function from higher to lower LoD. 
A hypothetical topological RDBMS---which might also become a basis for some database-backed CAD 
system in the future---could have topological data types $(X,R)$ instead of tables $X$ and 
should then provide topological DDL-statements like 
\begin{verbatim}
 CREATE SPACE LoD4
 ( spaceid INTEGER NOT NULL,
   id INTEGER NOT NULL,
   lod3space INTEGER,
   lod3id INTEGER,
   yaAttribute YATYPE, 
   ...
   PRIMARY KEY(spaceid, id),
   CONTINUOUS FOREIGN KEY
     (lod3space,lod3id) 
     REFERENCES LoD3(spaceid,id),
   ...);
\end{verbatim}
with the \texttt{CONTINUOUS} consistency constraint on foreign keys. 
A foreign key from one relation $R$ into another relation $S$ defines a function $f:R\to S$ 
that maps a referring tuple $r\in R$ to a referred tuple $f(r)\in S$. 
When $R$ and $S$ are spaces the database administrator could then establish a continuity 
restriction on $f$ whenever he needs it. The later example of a topologically integrated 
detail library will use this continuity constraint.

\subsection{Constructions and queries}

Resuming the modelling steps shows that, first, there was a \qq{selection} of some 
points from the embedding topological space which returned another topological space, 
the subspace. Then the selected points have been mapped into a set of features by a 
\qq{projection} which also not only converted points into features but additionally produced 
a topology for them. So these are already two analogs for relational query operators which 
leads to the question how relational algebra could act on a CAD model. 

A relationally complete set of query operations for topological data types 
can be found in \cite{BradleyPaul}. 
Table~\ref{tab:queries} lists some basic operators of relational algebra, their 
corresponding topological constructions, and the functions involved. 
Each query operator has functions that connect the input tables with the result table. 
These functions define the query result topology either as a \qq{final} topology when the 
functions map to the result, or as an \qq{initial} topology when the functions map the from 
the result to the input. 
It is always the topology that most resembles all input topologies but still keeps all 
linking functions continuous. A first experimental prototype of a topological relational 
algebra implemented by the author can be found on \url{http://pavel.gik.kit.edu}.  

\begin{table}[h]
  \centering
  \begin{tabular}{lll}
    Operation  & Construction & Function(s) \\
    \hline
    selection  & subspace    & $i: \sigma_\Theta(X) \hookrightarrow Xs$\\
    [\jot]
    projection & final space & $\pi_{\mathcal{A}}:X \to \pi_{\mathcal{A}}(X)$\\
    [\jot]
    union      & pasting     & $\check\imath_X:X \hookrightarrow X\cup Y$\\
               &             & $\check\imath_Y:Y \hookrightarrow X\cup Y$\\
    [\jot]
    intersection& pullback   & $\hat\imath_X: X\cap Y \hookrightarrow X $\\
               &             & $\hat\imath_Y: X\cap Y \hookrightarrow Y$\\
    [\jot]
    Cartesian product 
               & product space 
                             & $\pi_X: X\times Y \to X$\\
               &             & $\pi_Y: X\times Y \to Y$
  \end{tabular}
  \caption{Relational query operators, their corresponding topological 
constructions, and the continuous functions involved. The functions are either 
$f:X \to \texttt{op}(X,\dotsc)$ from a query input table $X$ to the query output table 
$\texttt{op}(X,\dotsc)$ or $g:\texttt{op}(X,\dotsc) \to X$ from the output 
to an input. When that arrow points to the output, the result topology 
is called a \qq{final} topology, otherwise it is called \qq{initial}. 
$i$, $\check\imath$, and $\hat\imath$ are inclusion functions.}
  \label{tab:queries}
\end{table}

The following discussion of Cartesian Product and join illustrates 
how topological constructions act on the relational representation of 
topological spaces.

\subsubsection{Cartesian product and join} 
The topological analog for the Cartesian product is the product space with 
its important practical application in CAD that it topologically generalises 
\emph{extrusion}. 
Let $(X,R)$ and $(Y,S)$ be two spatial data types, say, an axis $X$ and a profile $Y$. 
Then the Cartesian product is the set $X\times Y$ of entity-pairs $ts$ where $t$ is in $X$ 
and $s$ is in $Y$. 
The functions that link input with output are the \emph{two} projections 
$\pi_X:X\times Y\to X, ts \mapsto t$ and $\pi_Y:X\times Y\to Y, ts \mapsto s$. 
A maximal relation for $X\times Y$ that makes both projections continuous is made up of two 
components: 
A \qq{horizontal} part $\mathit{RY}$ that creates $Y$ copies of $R$ each \qq{tagged} by an element 
of $Y$ and a \qq{vertical} part $\mathit{XS}$ that creates $X$ copies of $S$ each \qq{tagged} by 
an element of $X$. 
The vertical part will be defined as 
$$
  \mathit{XS} := \mathpset{(xa, xb)}{x\in X, a\,S\,b}\enspace.
$$
Accordingly the \qq{horizontal} part is 
$$
  \mathit{RY} := \mathpset{(cy, dy)}{c\,R\,d, y\in Y}\enspace.
$$
The union $\mathit{XS} \cup \mathit{RY}$ is a binary relation for $X\times Y$ that 
creates the product space topology \cite{BradleyPaul}. 
So the Cartesian product space is generated by 
$$
  (X,R)\times(Y,S) := (X\times Y,\, \mathit{XS} \cup \mathit{RY})\enspace.
$$
Besides, it shall be mentioned that the topological dimension of spatial data 
(also called \qq{Krull dimension} \cite[p.\ 5]{Hartshorne:AlgebraicGeometry}) 
of the product space is the sum of the dimensions of the input spaces. 
The Cartesian product space of, say, two instances of a 3D data model 
will result in a 6D product space and would not fit into the restricted data model 
from Figure~\ref{fig:spatialUML}. 

A relational $\Theta$-join can now be formally defined as the subspace obtained by selecting 
the entity-pairs which satisfy the join-predicate $\Theta$ from the product space 
$$
  (X,R)\Join_\Theta(Y,S) := ((X,R)\times(Y,S))|_{\sigma_{\Theta}(X\times Y)}
$$
using the selection operator $\sigma$.

\section{More applications of continuity}\label{sec:applications2}

After having seen how continuous functions can define basic query operators 
two not-so-basic applications of topological constructions will be presented here.

\subsection{A topological detail library} 

The following construction is called \emph{fibre product} in topology and corresponds 
with the relational equi-join. 
The LoD-sequence of spaces---each a model of the same real-world object at a 
different level of detail---is linked by continuous functions \qq{part\_of} where each 
maps a finer object to a coarser representation. 
But engineering design often starts at the coarse side with sketchy design ideas that are 
later refined. 
It is also common in CAD-systems to define complex \qq{macro}-objects of 
frequently reused parts to be placed into a drawing. 
The following topological construction can do both: 

\begin{sloppy} 
\begin{dfn}[Fibre product]
Let $(X,\mathcal{T}_X)$, $(Y,\mathcal{T}_Y)$, and $(I,\mathcal{T}_I)$ be topological spaces and 
$f:(X,\mathcal{T}_X)\to(I,\mathcal{T}_I)$  and $g:(Y,\mathcal{T}_Y)\to(I,\mathcal{T}_I)$ be two 
continuous functions. 
Then the space 
$$
  (X,\mathcal{T}_X) \times_I (Y,\mathcal{T}_Y) := 
   \sigma_{f=g}((X,\mathcal{T}_X) \times (Y,\mathcal{T}_Y))
$$
is called the \emph{fibre product space} of $f$ and $g$. $\sigma_P(S)$ is the selection 
$\mathpset{s \in S}{s \text{ satisfies } P}$.
\end{dfn}
\end{sloppy} 

It is the space of all entity-pairs $xy$ from $X\times Y$ where $f(x)= g(y)$ holds. 
When $X$ and $Y$ are database tables with attributes $f$ and $g$ then $X \times_I Y$ is the 
equi-join $X \Join_{f=g} Y$.

The use of macro objects in a CAD drawing can be considered a fibre product: 
The set of the macro-names is a discrete space $I$ of identifiers.  
The locations in a drawing where such macro objects are used is another discrete space $X$. 
The use of macros establishes a function $f$ from the locations to the identifiers. 
This function is continuous because of its discrete domain. 
The set of all macros $Y$ is a so-called sum-space: a disjoint sum of topological spaces---one 
subspace per macro---each indexed by its macro identifier. 
Also the function $g$ that maps macro elements to macro identifiers is continuous if 
there is no connectivity between two different macros. 
Now the insertion of copies of macros into the drawing corresponds to the fibre product of the 
two continuous functions \qq{macro-use} $u$ and \qq{macro naming} $p$ which have a common range 
space, the discrete space $I$ of macro identifiers. 

Until now this was just a formal overkill but when the macro names get a topology and 
the macro objects space is coarsened, this turns into a system of a topologically integrated 
library of parts that are chosen and linked together by a coarser sketchy drawing and both 
together produce the detailed plan. 

\begin{figure}[t]
\begin{center}
\includegraphics[width=\columnwidth]{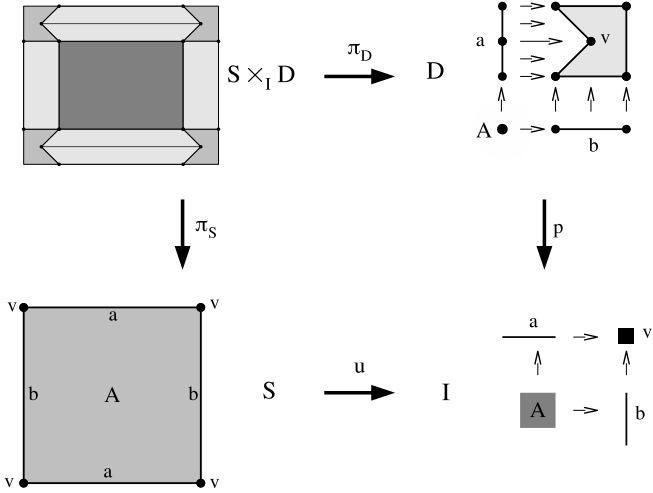}
\end{center}
\caption{The fibre product space $S\times_I D$ of the continuous functions ${p:D\to I}$ and 
${u:S\to I}$ (the detail \textbf{\textit{u}}sage) \qq{topologically integrates} the space of 
details $D$, the space of the detail index $I$ and the sketch $S$ into a more detailed space.}
\label{fig:FiberProduct}
\end{figure}

The following is a simple 2D-example but the intended application is 3D CAD:
Figure~\ref{fig:FiberProduct} shows a sketchy drawing $S$ on the lower left hand side and a 
space of some details $D$ on the upper right hand side. Both are topological spaces with 
continuous functions $u:S\to I$ and $p:D\to I$ that map into a common index space $I$. 
$u$ stands for \qq{usage} and denotes which detail from the detail library $p:D\to I$ shall be 
used. 
The \qq{usage}-values of $u$ are depicted in $S$: The area element \qq{uses} $A$, the $u$-value 
of the two horizontal lines is $a$, that of the vertical lines it is $b$, and so on. 
The same is true for the $p$-values in $D$. 
The little arrows in $D$ and $I$ indicate the topological incidence relation \emph{between} 
the single detail entities. Therefore the details are no longer a disjoint sum.  
For example, detail $a$ consists of five entities of which the middle vertex 
is \qq{close} to the vertex in the notch of detail $v$. 

The detail of name $x\in I$ is the subspace of the elements in $D$ that map to 
$x$, hence the pre-image $p^{-1}(x)$ of $x$ under $p$. 
Pre-images of points are also called \qq{fibres} and therefore this construction is named 
\qq{fibre product}. For example, the detail $a$ is the sequence of the three vertices 
connected by two edges in the upper left corner of $D$. 

The \emph{fibre product space} $S \times_I D$ is generated be these two functions: 
It is the topological equi-join $S\Join_{u=p} D$ depicted in the upper 
left corner of Figure~\ref{fig:FiberProduct}. For example, the upper horizontal \qq{wall} is 
an extrusion of the detail $a$ along the upper horizontal edge and it seamlessly connects 
to the notch of the vertex detail $v$. 
This operation has some interesting features for 3D CAD and CAE design: 
\begin{itemize}
\item The details are a topological generalisation of \qq{macro} objects in CAD. 
\item It is possible to specify the connectivity between details in a uniform manner. 
\item The detail library, the continuous function $p$, establishes consistency rules on how 
      details can be used and which details can be \qq{compatibly} connected. 
\item A user working on the drawing $S$ can only select his details from $I$, as long as his 
      \qq{detail use}-function $u$ remains continuous. The continuity constraint prohibits 
      incompatible connections and thereby again advocates for continuity as a consistency rule. 
\item Details can be placed at singular locations like traditional macro objects but can also be 
      \qq{extruded} along extended objects: 
      A two-dimensional profile can be extruded along a one-dimensional axis whereas a 
      one-dimensional sequence of wall-layers can be extruded along a 2D wall 
      face to generate a 3D-wall sandwich element. All these extrusions generate 3D-objects 
      and the components' dimensions sum up to 3. 
\end{itemize}

There are still open questions: For example, a naive use of the pullback can return a topology 
that is too coarse for practical applications, hence, some \qq{refinement} of the concept is still 
necessary. But it surely will be the essential ingredient of any topologically integrated detail 
library application for CAD systems.

\subsection{Intersection} 

Until now only combinatorial topological operations have been considered. 
The computation of the intersection space that results in overlaying 
geometrically embedded spatial data is another important operation. 
In GIS, for example, this \qq{overlay} is often used to carry out combined spatial 
data analysis. 

Let there be a predicate $\Theta$ that tells for two given elements $a$ and $b$, each from 
one space, if they geometrically intersect. 
Iff they intersect $\Theta(a,b)$ is true. An algorithm for $\Theta$ on $n$D-polytope complexes 
is sketched in \cite{SimplicialDecomposition}. 

\begin{figure}
\begin{center}
\includegraphics[width=\columnwidth]{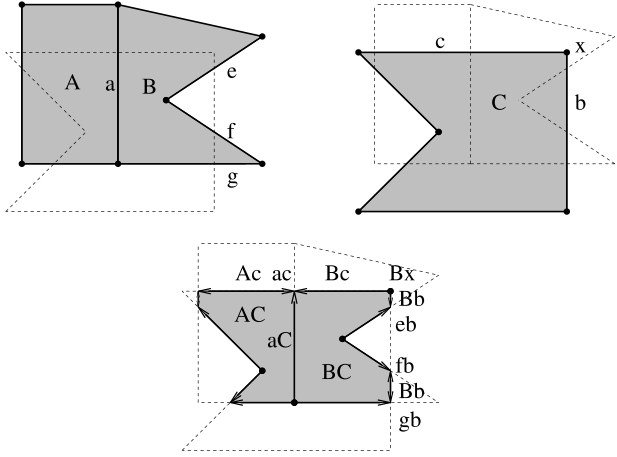}
\end{center}
\caption{The intersection space of two spaces of geometrically intersecting entities. 
Topologically it is the two-dimensional subspace of intersecting cell pairs within the 
four-dimensional product space. 
The combinatorial \qq{close-to} relation computed by the relational query 
operators indeed matches the metric \qq{close-to} from the embedding of the intersection. 
The two projections from the intersection back to the intersecting spaces are continuous.}
\label{fig:Intersection}
\end{figure}

With two topological data types $(X,R)$ and $(Y,S)$ and such an \qq{intersects} predicate 
$\Theta$ the intersection space is the $\Theta$-Join 
$$
  (X,R)\Join_{\Theta}(Y,S) := \sigma_{\Theta}((X,R)\times(Y,S))\enspace.
$$
It consists of all pairs of elements that have a geometrical intersection. 
Clearly, the intermediate Cartesian product space computation is expensive and should be 
avoided in practice. Codd calls this emphasis of avoiding the 
Cartesian product wherever possible its \qq{de-emphasising} \cite{Codd:RMV2}. 
Similar to the fibre product the join also gets a topology for these pairs. 
Figure~\ref{fig:Intersection} gives an example. 
Note that by explicitly storing $\Theta$ into an association $\Theta\subseteq X\times Y$ the three 
items $(X,R)$, $(Y,S)$, and $\Theta$ realise the $Q$-Complex proposed in \cite{Qomplex}. 

Some elements from the two intersected spaces in Figure~\ref{fig:Intersection} 
show how the $\Theta$-join computes the topology. 
The left-hand side space contains two faces $A$ and $B$, eight edges---among which four are 
labelled $a$, $e$, $f$, and $g$---and seven vertices.  
The right-hand side space is made of only one face $C$ bounded by five edges where two are 
labelled $b$ and $c$, and five vertices among one is named $x$. 
The incidence relations for the named entities are 
$$
 R = \mathset{(A,a),(B,a),(B,e),(B,f),(B,g)}
$$
and 
$$
 S = \mathset{(C,c),(C,b),(c,x),(b,x)}\enspace.
$$
The pair $(C,x)$, which denotes that vertex $x$ is close to face $C$, is in $S^{+}$, the 
transitive closure of $S$. 
The intersections of the elements in $X\supset\mathset{A,B,a,e,f,g}$ with those in 
$Y\supset\mathset{C,b,c,x}$ can be decided by feeding each pair into $\Theta$. 
For example, $\Theta(A,c)$ returns \texttt{true} and $\Theta(A,b)$ returns \texttt{false}. 
The following 24-intersections matrix depicts all pairs of intersecting named entities,  
that is, all $pq$ where $\Theta(p,q)$ is \texttt{true}. 
\begin{center}
\begin{tabular}{c|cccccc}
   & $A$& $B$& $a$& $e$& $f$& $g$\\\hline
$C$&$AC$&$BC$&$aC$&$eC$&$fC$&$gC$\\
$b$&    &$Bb$&    &$eb$&$fb$&$gb$\\
$c$&$Ac$&$Bc$&$ac$&    &    &    \\
$x$&    &$Bx$&    &    &    &    
\end{tabular}
\end{center}
In the product space the pair $aC$ is \qq{close to} $AC$ because $a$ is close to $A$ 
and for all elements $q$ in the other space the pair $(A,a)$ in $R$ is converted into $(Aq,aq)$. 
As $C$ is one such $q$ the incidence $(AC,aC)$ belongs to the incidence relation 
of the product space. 
Conversely, the pair $(C,c)$ in the right-hand side incidence relation has been turned into 
the incidences 
\begin{align*}
  \{&(AC,Ac), (BC,Bc), (aC,ac),\\
    &(eC,ec), (fC,fc), (gC,gc),\dotsc\}
\end{align*}
telling, for example, that $Bc$ is close to $BC$. 
From that four-di\-men\-sio\-nal product space the predicate $\Theta$ selects the intersecting 
entity pairs. Together with its subspace incidence relation this gives a two-dimensional result 
space. 
The projections from the intersection back into the input are continuous functions: 
The left hand side projection takes, for example, the incident elements $AC$ and $Ac$ to $A$ 
and so by ${\pi_X(AC)=A=\pi_X(Ac)}$ this pair satisfies the continuity condition for the 
projection \newline
${\pi_X:X\Join_{\Theta}Y\to X}$. 
The right hand side projection takes $AC$ to $C$ and $Ac$ to $c$ and so 
$(\pi_Y(AC),\pi_Y(Ac)) = (C,c) \in S$. Therefore the pair also satisfies the continuity 
condition for the other projection 
$\pi_Y:X\Join_{\Theta}Y\to Y$. 

Interestingly the edge $Bb$ has four vertices $Bx$, $eb$, $fb$, and $gb$. So $Bb$ 
became an edge with a hole. This is how the $\Theta$-join produces this boundary of $Bb$: 
The involved entities are $B$, $e$, $f$, and $g$ in $(X,R)$ and $b$, and $x$ in $(Y,S)$. 
The involved incidence relations are 
\begin{align*}
  \mathset{(B,e),(B,f),(B,g)}&\subset R \\
  \text{~and~}\mathset{(C,b),(b,x)}&\subset S\enspace.
\end{align*}
The following 17-products-matrix shows the incidence relation for the Cartesian product space of 
the subspace near $Bb$. An entry which depicts a non-intersection is \sout{stroked out} and 
pairs where $Bb$ is at the left-hand side are highlighted in \textbf{bold face}:
\begin{center}
\newcommand{\so}[1]{\sout{#1}}
\newcommand{\bi}[1]{\textbf{\textit{#1}}}
\it
\begin{tabular}{@{}c@{~}|@{~}c@{~}c@{~}c@{~}|@{~}c@{~}c@{}}
       &    C     &       b       &       x      &  (C, b)  &    (b, x)     \\\hline
 B     &          &               &              & (BC, Bb) & \bi{(Bb, Bx)} \\[\jot] 
 e     &          &               &              & (eC, eb) & \so{(eb, ex)} \\[\jot] 
 f     &          &               &              & (fC, fb) & \so{(fb, fx)} \\[\jot] 
 g     &          &               &              & (gC, gb) & \so{(gb, gx)} \\[\jot] \hline
(B, e) & (BC, eC) & \bi{(Bb, eb)} & \so{(Bx, ex)} \\[\jot] 
(B, f) & (BC, fC) & \bi{(Bb, fb)} & \so{(Bx, fx)} \\[\jot] 
(B, g) & (BC, gC) & \bi{(Bb, gb)} & \so{(Bx, gx)} 
\end{tabular}
\end{center}
This gives the incidences $(Bb,Bx)$, $(Bb,eb)$, $(Bb,fb)$, and $(Bb,gb)$, 
hence $Bx$, $eb$, $fb$, and $gb$ are in the boundary of $Bb$. They are 
vertices because they do not occur as a left-hand side element in an incidence and 
having an empty boundary characterises vertices in the topological model. 
Note that $eb$ is a zero-dimensional vertex in the $\Theta$-join space 
but it was a two-dimensional element in the intermediate product space because it is the pair  
of the two edges $e$ and $b$, each of dimension 1, and in the product space the dimensions of 
elements add. 
This shows how the generic concept of a topological data type allows to dynamically shift the 
dimension upper bound up and down as necessity arises.

\section{Conclusion and outlook}

The research which r\^ole continuous functions might play in CAD data models has 
shown that they link the embedding space with the spatial data. 
These data models are topological spaces, and so continuity is well-defined for spatial data. 
It has been demonstrated how continuity of functions can be a versatile tool for spatial 
data modelling and can be used as a consistency rule. 
Continuity also suggests relaxing the current strict-typed topological models with their 
classical volume-face-edge-vertex sequences and have a closer look at their common generalisation 
\texttt{SpatialObject}. 

Topological constructions are based on continuity and it has been shown that all basic query 
operators of relational algebra have corresponding topological query operators for spatial data. 
So the relational representation not only enormously simplifies spatial data modelling but 
also adds expressive power to spatial data management. 
This topological relational algebra could be part of a framework for future database-backed CAD 
systems. In particular, the fibre product has been identified as an extremely promising 
topological construction for CAD and CAE applications. 

To wrap it up, it has been shown that the importance of continuous functions in topology 
reflects onto CAD modelling, and the potentials of mathematical topology have been demonstrated 
here to go far beyond giving predicate names to intersection patterns. 
In addition to providing a theoretical framework and to its relationally complete set of 
query operators, topology still offers many concepts that just wait for being harvested 
from literature and used in future CAD and CAE applications.

\section{Acknowledgements}

This work was funded by the German Research Foundation (Deutsche Forschungsgemeinschaft DFG) with 
research grant BR2128/12-2. The author thanks Patrick Erik Bradley for valuable remarks and 
discussions on this topic. 

\bibliographystyle{elsarticle-num}
\bibliography{biblio}









\end{document}